\documentclass[graybox]{svmult}

\usepackage{amsmath,amssymb,amsfonts}
\usepackage{mathptmx}       % selects Times Roman as basic font
\usepackage{helvet}         % selects Helvetica as sans-serif font
\usepackage{courier}        % selects Courier as typewriter font
\usepackage{type1cm}        % activate if the above 3 fonts are
\usepackage{makeidx}         % allows index generation
\usepackage{graphicx}        % standard LaTeX graphics tool
\usepackage{multicol}        % used for the two-column index
\usepackage[bottom]{footmisc}% places footnotes at page bottom
\usepackage[numbers]{natbib}
\usepackage{comment}

\makeatletter
\providecommand{\leftsquigarrow}{%
  \mathrel{\mathpalette\reflect@squig\relax}%
}
\newcommand{\reflect@squig}[2]{%
  \reflectbox{$\m@th#1\rightsquigarrow$}%
}
\makeatother

\newcommand{\leadsfrom}{\leftsquigarrow}

  \usepackage{tikz}
\usetikzlibrary{snakes}
\usetikzlibrary {shapes}
\usetikzlibrary {arrows}
\usetikzlibrary {positioning}
\usetikzlibrary {calc}
\usetikzlibrary{fit}					% fitting shapes to coordinates
\usetikzlibrary{backgrounds}	% drawing the background after the foreground

\makeindex             % used for the subject index
\newcommand{\bthe}{\begin{theorem}}
\newcommand{\ethe}{\end{theorem}}

\newcommand{\ben}{\begin{enumerate}}
\newcommand{\een}{\end{enumerate}}

\newcommand{\bit}{\begin{itemize}}
\newcommand{\eit}{\end{itemize}}

\newcommand{\beq}{\begin{equation}}
\newcommand{\eeq}{\end{equation}}

\newcommand{\ble}{\begin{lemma}}
\newcommand{\ele}{\end{lemma}}

\newcommand{\bde}{\begin{definition}\rm}
\newcommand{\ede}{\halmos\end{definition}}

\newcommand{\bco}{\begin{corollary}}
\newcommand{\eco}{\end{corollary}}

\newcommand{\bpr}{\begin{proposition}}
\newcommand{\epr}{\end{proposition}}

\newcommand{\brem}{\begin{remark}\rm}
\newcommand{\erem}{\halmos\end{remark}}

\newcommand{\bproof}{\begin{proof}}
\newcommand{\eproof}{\halmos\end{proof}}

\newcommand{\bexam}{\begin{example}\rm}
\newcommand{\eexam}{\halmos\end{example}}

\newcommand{\bfi}{\begin{fig}}
\newcommand{\efi}{\end{fig}}

\newcommand{\btab}{\begin{tab}}
\newcommand{\etab}{\end{tab}}

\newcommand{\beao}{\begin{eqnarray*}}
\newcommand{\eeao}{\end{eqnarray*}\noindent}

\newcommand{\beam}{\begin{eqnarray}}
\newcommand{\eeam}{\end{eqnarray}\noindent}

\newcommand{\barr}{\begin{array}}
\newcommand{\earr}{\end{array}}

\def\R{{\mathbb R}}

\def\calc{{\mathcal{C}}}

\def\calg{{\mathcal{G}}}
\def\calp{{\mathcal{P}}}

\def\calx{{\mathcal{X}}}

\newcommand{\graf}{\calg}
\newcommand{\coeff}{C}
\newcommand{\Coeff}{\calc}

\def\1{\mathbf{1}}
\newcommand{\dd}{\mathrm{d}}

\newcommand\ci{{\perp\!\!\!\!\perp}}
\newcommand{\cip}{\ci}
\newcommand{\gse}{\perp_\calg}
\newcommand{\dse}{\perp_\dag}
\newcommand{\mse}{\perp_m}
\newcommand{\sg}{{\perp_\sigma}}
\newcommand{\cd}{\,|\,}

\newcommand{\spa}{\calx}

\newcommand{\la}{\lambda}

\newcommand{\DAG}{{\rm DAG}}
\renewcommand{\dag}{\mathcal{D}}

\newcommand{\sgn}{{\rm sgn}}
\newcommand{\supp}{{\rm supp}}

\newcommand{\an}{{\rm an}}
\newcommand{\pa}{{\rm pa}}
\newcommand{\parents}{\pa}
\newcommand{\pr}{{\rm pr}}
\newcommand{\nd}{{\rm nd}}

\newcommand{\An}{{\rm An}}
\newcommand{\Pa}{{\rm Pa}}

\newcommand{\Des}{{\rm De}}

\newcommand{\ch}{{\rm ch}}
\newcommand{\ske}{{\rm ske}}

\newcommand{\halmos}{\quad\hfill\mbox{$\Box$}}  

\newcommand{\CK}[1]{{\color{blue} #1}}
\newcommand{\SL}[1]{{\color{red}#1}}

\begin{document}

\title*{Bayesian Networks for Max-linear Models}
\author{Claudia Kl\"uppelberg and Steffen Lauritzen}
\institute{Claudia Kl\"uppelberg \at  Center for Mathematical Sciences, Technical University of Munich,  85748 Garching, Boltzmannstrasse 3, Germany; \email{cklu@tum.de}
\and Steffen Lauritzen \at Department of Mathematical Sciences, University of Copenhagen, Universitetsparken 5, 2100 Copenhagen, Denmark; \email{lauritzen@math.ku.dk}}
\maketitle

\abstract*{Each chapter should be preceded by an abstract (10--15 lines long) that summarizes the content. The abstract will appear \textit{online} at \url{www.SpringerLink.com} and be available with unrestricted access. This allows unregistered users to read the abstract as a teaser for the complete chapter. As a general rule the abstracts will not appear in the printed version of your book unless it is the style of your particular book or that of the series to which your book belongs.
Please use the 'starred' version of the new Springer \texttt{abstract} command for typesetting the text of the online abstracts (cf. source file of this chapter template \texttt{abstract}) and include them with the source files of your manuscript. Use the plain \texttt{abstract} command if the abstract is also to appear in the printed version of the book.}

\abstract{We study Bayesian networks based on max-linear structural equations as introduced in \citet{GK1} and provide a summary of their independence properties. In particular we emphasize that distributions for such networks are generally not faithful to the independence model determined by their associated directed acyclic graph. 
In addition, we consider some of the basic issues of estimation and discuss generalized maximum likelihood estimation of the coefficients, using the concept of a generalized likelihood ratio for non-dominated families as introduced by \citet{KW1956}. Finally we argue that the structure of a minimal network asymptotically can be identified completely from observational data.}

\section{Introduction}

The type of model we are studying has been motivated by applications to risk analysis,
 where extreme risks play an essential role and may propagate through a network. For example, say, if an extreme rainfall happens on a specific location near a river network, it may effect water levels at other parts of the network in an essentially deterministic fashion.  Similar phenomena occur in the analysis of risk for other complex systems.
 
Specifically, the model presented in \eqref{ml-sem1} below arose in the context of technical risk analysis, more precisely, in an investigation of the ``runway overrun'' event of airplane landing. 
Numerous variables contribute to this event and extraordinary values of some variables lead invariably to a runway overrun (see \cite{GKM} for more details) naturally leading to questions about cause and effect of risky events.
Other potential examples for risk-related cause and effect relations include chemical pollution of rivers (\cite{HPT}), flooding in river networks (\cite{Asadi2015}), financial risk (\cite{EKS}), and many others. 

Statistical theory and applications of extreme value theory until the 1990s mainly focused on i.i.d.\ data as, for instance, yearly maximal water levels to predict future floodings or peaks over thresholds used to estimate the Value-at-Risk (e.g. \cite{EKM}).
From this, both theory and applications moved on to multivariate data, modelling risks like joint wind and wave extremes as well as extreme risks in financial portfolios \cite{Beirlant2004}.
The investigation of extremes in time series models have proved useful in financial and environmental risk analysis, and also in telecommunication (see e.g. the book \cite{FR}).
More recently, extreme space-time models have been suggested and applied to environmental risk data \cite{BDKS,DKS,DavisonPadoanRibatet,Huser}.

The paper focuses on first steps reporting on the methodological development associated with  a specific class of network models. 
We begin with introducing our leading example of a recursive max-linear model which is Example 2.1 of \cite{GK1}:
\bexam\label{ex:maxlin}
Consider the network in the figure below:
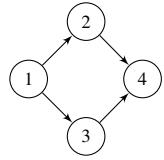
\begin{center}
\begin{tikzpicture}[node distance = 4mm and 4mm, minimum width = 5mm]
    %% nodes      
    \begin{scope}
      \tikzstyle{every node} = [shape = circle, 
      font = \scriptsize,
      minimum height = 4mm,
      inner sep = 0pt,
      draw = black, 
      fill = white, 
      anchor = center], 
      text centered] 
      \node(a) at (0,0) {$1$};
      \node(b) [above right = of a] {$2$};
      \node(c) [below right = of a] {$3$}; 
      \node(d) [right =10mm of a]{$4$};
    \end{scope}  

    %% directed edges
    \begin{scope}[->, > = latex']
		\draw (a) -- (b);
    \draw (a) -- (c);
   \draw (c) -- (d);
	\draw (b) -- (d);
    \end{scope}

    \end{tikzpicture}
    \end{center}
   
   Each node $i$ in the network represents a random variable $X_i$ and the joint distribution of $X=(X_1,X_2,X_3,X_4)$ is determined by a system of \emph{max-linear structural equations}
  \[X_1=Z_1,\; X_2=\max(c_{21}X_1,Z_2),\; X_3=\max(c_{31}X_1,Z_3),\; \max(c_{42}X_2,c_{43}X_3,Z_4),\]
  where $Z_1,Z_2,Z_3,Z_4$ are independent positive random variables and the coefficients $c_{ji}$ are all strictly positive.

The interpretation of a system like this is that each node in the network is subjected to a random shock $Z_i$ and  the effect from shocks of other nodes pointing to it, the latter being attenuated or amplified by the coefficients $c_{ji}$.  To simplify notation here and later we write
$a\vee b$ for $\max(a,b)$. We can alternatively represent $X=(X_1,X_2,X_3,X_4)$ directly in terms of the noise variables as
\begin{align*}
X_1 &=
Z_1\\
X_2 &= c_{21} X_1\vee Z_2 = c_{21} Z_1\vee  Z_2\\
X_3 &= c_{31} X_1\vee Z_3 = c_{31} Z_1 \vee   Z_3\\
X_4 &= c_{42} X_2 \vee c_{43} X_3\vee  Z_4 \\
&=c_{42} (c_{21} Z_1\vee Z_2) \vee c_{43}(c_{31}Z_1 \vee Z_3)\vee  Z_4\\
&=  ( c_{42}c_{21}\vee c_{43}c_{31}) Z_1 \vee c_{42}  Z_2 \vee c_{43} Z_3\vee  Z_4.
\end{align*}
We may then summarize the above coefficients to the noise variables $Z_1,\dots,Z_4$ in the matrix
\begin{align*}
B = 
\begin{pmatrix}
1&  0    & 0 &  0   \\
c_{21}  & 1&   0    &  0 \\
c_{31}    &    0   &   1&   0  \\
c_{42}c_{21} \vee c_{43} c_{31}& c_{42} & c_{43} &   1
\end{pmatrix},
\end{align*}
\eexam In greater generality we may write such a \emph{recursive max-linear model} as
\begin{align} \label{ml-sem1}
X_v =\bigvee\limits_{u \in \pa(v)} c_{vu} X_k \vee c_{vv} Z_v,\quad v=1, \dots, d,
\end{align}
where $\pa(v)$ denotes parents of $v$ in a directed acyclic graph (DAG) and $Z_v$ represent independent noise variables. The present article is concerned with such models and summarizes basic elements of \citet{GK1} and \citet{GKL}.

In this setting, natural candidates for the noise distributions are extreme value distributions or distributions in their domains of attraction resulting in a corresponding multivariate distribution with dependence structure given by the \DAG\ (for details and background on multivariate extreme value models see e.g. \cite{DHF,Resnick1987,Resnick2007}). 
The paper is structured as follows.

In Section~\ref{sec:prelim} we establish the necessary terminology (Section~\ref{sec:graph}), introduce Bayesian networks (Section~\ref{sec:Bayes}), and basic properties of conditional independence (Section~\ref{sec:cond}). In Section~\ref{sec:bn} we establish basic Markov properties of Bayesian networks. In Section~\ref{sec:rec} we study the specific Markov properties of Bayesian networks given by max-linear structural equations as in \eqref{ml-sem1} and in Section~\ref{sec:stats} we study statistical properties of the models.

\section{Preliminaries}\label{sec:prelim}

\subsection{Graph terminology}\label{sec:graph}

A \emph{graph}  as we use it here is determined by a finite vertex set $V$, an edge set $E$, and a map that to each edge $e$ in $E$ associates its endpoints $u, v\in V$. Our graphs are \emph{simple} so that there are no self-loops (edges with identical endpoints) and no multiple edges. Therefore we can identify an edge $e$ with its endpoints $u,v$ so we can write $e=uv$. An edge $uv$  of a \emph{directed} graph points \emph{from} $u$ \emph{to} $v$ and we write $u\to v$. Then $u$ is a \emph{parent} of $v$ and $v$ is a \emph{child} of $u$. The set of parents of $v$ is denoted $\pa(v)$ and the set of children of $u$ is $\ch(u)$.
If $uv$ is an edge we also say that $u$ and $v$ are adjacent and write $u\sim v$ whether or not the edge is directed.

 A \emph{walk} $\omega$ from $u$ to $v$ of \emph{length} $n$ is a sequence of vertices $\omega =[u=u_0, u_1, \ldots, u_n=v]$ so that $u_{i-1}\sim u_{i}$ for all $i=1,\ldots,n$.  A walk is a \emph{cycle} if $u=v$. A \emph{path} is a walk with no repeated vertices. 
 The walk is \emph{directed} from $u$ to $v$ if $u_{i-1}\to u_{i}$ for all $i$.  If all edges in a graph $\dag=(V,E)$ are directed, $\dag$ is a \emph{directed graph}. A directed graph is \emph{acyclic} if it has no directed cycles. A \emph{DAG} is  a directed acyclic graph.   A DAG is a \emph{tree} if every vertex has at most one parent and a \emph{polytree} if there is at most one path between two vertices $u$ and $v$.

 If there is a directed path from $u$ to $v$ in $\dag$ we say that $u$ is an \emph{ancestor} of $v$ and $v$ a \emph{descendant} of $u$ and write $u\leadsto v$ or $v\leadsfrom u$. The set of ancestors of $v$ is denoted $\an(v)$. A set $A\subseteq V$ is said to be \emph{ancestral} if $\an(v)\subset A$ for all $v\in A$, or, alternatively, if $\pa (v)\subset A$ for all $v\in A$. For a subset $A$ of $V$ we let $\An(A)$ denote the smallest ancestral set containing $A$.   
 
 We say that the vertex set $V$ of a DAG $\dag$ is \emph{well-ordered} if $V=\{1,\ldots,d\}$ and all edges in $\dag$ point from low to high, i.e.\ if $ij\in E \implies i<j$. Then the set of \emph{predecessors} of  a vertex $i$ is  $\pr(i) =\{1,\ldots, i-1\}$.

\begin{figure}[htb]
\centering
 \begin{tikzpicture}[node distance = 4mm and 4mm]
    %% nodes      
    \begin{scope}
      \tikzstyle{every node} = [shape = circle, 
      font = \scriptsize,
      minimum height = 4mm,
      minimum width = 4mm,
      inner sep = 0pt,
      draw = black, 
      fill = white, 
      anchor = center, 
      text centered] 
      \node(a) at (0,0) {$1$};
      \node(c) [right = of a] {$3$};
      \node(b) [below = of a] {$2$};
      \node(d) [below = of c] {$4$}; 
      \node(e) [right = of c] {$5$};
      \node(f) [right = of d] {$6$};
      
      \node(a1) [right = 10mm of e] {$1$};
      \node(c1) [right = of a1] {$3$};
      \node(b1) [below = of a1] {$2$};
      \node(d1) [below = of c1] {$4$}; 
      \node(e1) [right = of c1] {$5$};
      \node(f1) [right = of d1] {$6$};
      
    \end{scope}  
    %
    
    %% directed edges
    
    \begin{scope}[->, > = latex']
    \draw (a) -- (c);
    	\draw (b) -- (c);
    	\draw  (c) -- (e);
    	\draw (c) -- (e);
    	\draw (d) -- (f);
    	 \draw  (e)--(f);
    \end{scope}
    %
    
    %% undirected edges
    \begin{scope}
      \draw  (a1) -- (b1);
      %\draw  (c1)--(d1);
      \draw  (e1)--(f1);
      \draw  (e1)--(d1);
      \draw (a1) -- (c1);
    	\draw  (b1) -- (c1);
    	\draw (c1) -- (e1);
    	\draw (c1) -- (e1);
    	\draw (d1) -- (f1);
    \end{scope}
    %
  
    %% figure labels
    \begin{scope}
    \tikzstyle{every node} = [node distance = 4mm and 4mm, minimum width = 4mm,
    font= \scriptsize,
      anchor = center, 
      text centered] 
\node(a4) [below = 2mm of d]{$\dag$};
\node(b4) [below = 2mm of d1]{$\dag^m$};
\end{scope}
\end{tikzpicture}
 \caption{\label{fig:morchaingraph}
 A DAG $\dag$ and its moral graph $\dag^m$.  In $\dag$, $3$ has parents $1,2$ and $5$ is a child of $3$.  The DAG $\dag$ is a polytree. The node $6$ is a descendant of $1$, and $2$ is an ancestor of $4$. The set $\{1,2,3,5\}$ is ancestral in $\dag$. With the node numbering given, the DAG is well-ordered.
 }
\end{figure}
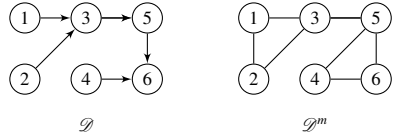

 For a DAG  $\dag$ we define   its {\em moral graph\/}  $\dag^{m}$  as the simple, undirected
graph with the same vertex set but with $u$ and
$v$ adjacent in $\dag^{m}$ if and only if either
$u\sim v$ in $\dag$ or if $u$ and $v$ have a common child.  For further general graph terminology we refer the reader to \citet{west:01} but  some of the concepts above are illustrated in Fig.~\ref{fig:morchaingraph}.

\subsection{Bayesian networks}\label{sec:Bayes}

A real-valued Bayesian network  associated to a given DAG $\dag=(V,E)$ is determined by specifying  random variables $X=(X_v, v\in V)$ and the conditional distribution of each of these, given values of their parent variables; for example as
\[P(X_v\leq x\cd X_{\pa(v)})=F(x\cd x_\pa(v)).\]
Because there are no directed cycles in $\dag$ there is a unique joint distribution corresponding to this specification. 

Alternatively, as in Example~\ref{ex:maxlin}, we can specify these conditional distributions through \emph{structural equations} which describe the conditional distribution of $X_v$ conditionally on $X_{\pa(v)}=x_{\pa(v)}$ in a functional form. More precisely a system of equations of the form
\begin{equation}\label{eq:structure}
X_v=g_v(X_{\pa(v)},Z_v),\quad  v \in V,
\end{equation}
where $(Z_v)_{v\in V}$ are independent noise variables and $g_v$ suitable functions.

A system of structural equations as above is sometimes referred to as a \emph{data generating mechanism}, interpreting each equation as a way of generating random variables with the desired conditional distribution. 

An important instance of these models are \emph{linear structural equation models} where the functions $g_v$ are linear and hence 
\begin{equation}\label{eq:linear}
X_v=\sum_{u\in \pa(v)}c_{vu}X_u+ c_{vv}Z_v, \quad  v \in V,\end{equation}
where $c_{vu}, u\in \pa(v), c_{vv}$ are \emph{structural coefficients}, see for example \citet{Bollen}. In general, a structural equation system need not be associated with a DAG, but if it is, the equation system is said to be \emph{recursive}.

If the distributions of $Z_v$ have heavy tails and all structural coefficients are non-negative, the sum tends to be dominated by the largest term:
\[\sum_{u\in \pa(v)}c_{vu}X_u+ c_{vv}Z_v\approx \bigvee_{u\in \pa(v)}c_{vu}X_u\vee c_{vv}Z_v\]
and hence for such cases, the max-linear variant in (\ref{ml-sem}) as described in more detail in Section~\ref{sec:rec} below.

\subsection{Conditional independence}\label{sec:cond}

The notion of conditional independence is at the heart of graphical models, including Bayesian networks. For three random variables $(X, Y, Z)$  we say that $X$ is conditionally independent of $Y$ given $Z$  if the conditional distribution of $X$ given $(Y, Z)$ does not depend on $Y$ and we then write $X\cip Y\cd Z$ or $X\ci_P Y\cd Z$ if we wish to emphasize the dependence on the joint distribution $P$ of $(X,Y,Z)$. 

The notion of conditional independence has a number of important properties, see e.g.\ \citet{dawid:80} or \citet{Lauritzen1996}.

\begin{proposition}\label{prop:ciproperties} Let $(\Omega,\mathbb{F},P)$ be a probability space and  $X$, $Y$, $Z$, $W$ random variables on $\Omega$. 
Then the following properties hold.
\begin{description}
\item[\rm (C1)]If $X\cip Y\cd Z$ then $Y\cip X\cd Z$ (symmetry);
\item[\rm (C2)]If $X\cip Y\cd Z$  and $W=\phi(Y)$ then $X\cip W\cd Z$ (reduction);
\item[\rm (C3)]If $X\cip (Y, Z)\cd W$  then $X\cip Y\cd (Z,W)$ (weak union);
\item[\rm (C4)]If $X\cip Z\cd Y$ and $X\cip W\cd (Y,Z)$ then $X\cip (Z,W)\cd Y$ (contraction);
\end{description}
\end{proposition}

It is occasionally important to abstract the notion of conditional independence away from necessarily being concerned with probability measures.  An (abstract) \emph{independence model} $\sg$ over $V$ is a ternary relation over subsets of a finite set $V$. The independence model is a \emph{semi-graphoid} if the following holds 
 for mutually disjoint subsets $A$, $B$, $C$, $D$:
\begin{description}
        \item[\rm (S1)] \emph{If $A \sg B \cd C$ then $B \sg A \cd C$ {(symmetry)};}
        \item[\rm (S2)] \emph{If $A \sg (B\cup D) \cd C$  then $A
\sg B\cd C$ and  $A\sg D\cd C$ {(decomposition)};}
 \item[\rm (S3)] \emph{If $A \sg (B \cup D) \cd C$ then $A \sg
   B \cd
(C\cup D)$ {(weak union)};}
 \item[\rm (S4)] \emph{If $A \sg B \cd C$ and $A \sg D \cd (B\cup C)$,
  then $A \sg (B\cup D) \cd C$ {(contraction)};}
  \end{description}
  Further, the
independence model is  a \emph{graphoid} if it also satisfies 
\begin{description}
        \item[\rm (S5)]  \emph{If $A \sg B \cd (C\cup D)$ and $A \sg C \cd
(B\cup D)$ then
$A \sg (B\cup C)\cd D$ (intersection).}
 \end{description}

We shall in particular be interested in distributions on product spaces $\calx=\times_{v\in V}\calx_v$ where $V$ is a finite set. For $A\subseteq V$ we write $x_A=(x_v, v\in A)$ to denote a generic element in$\calx_A=\times_{v\in A}\calx_v$, and similarly $X_A=(X_v)_{v\in A}$. 

If $P$ is a probability distribution on $\calx$, we can now define an independence model $\ci$ by the relation
\[A\ci B\cd C \iff X_A\ci_P X_B\cd X_C\]and it follows from Proposition~\ref{prop:ciproperties} that \emph{$\ci$ is a semi-graphoid}; in general $\ci$ is not a graphoid without further assumptions on $P$. 

Another important independence model is determined by \emph{separation} in an undirected graph. More precisely, if $\calg=(V,E)$ is an undirected graph we can define an independence model $\gse$ by letting $A\gse B\cd S$ mean that all paths in $\calg$ from $A$ to $B$ intersect $S$. Then it is easy to see that $\gse$ is always a graphoid; indeed the term graphoid refers to this fact. 

For a directed graph, the relevant notion of separation is more subtle. A vertex $u$ is a \emph{collider} on a path $\pi$ if two arrowheads meet on the walk at $u$, i.e.\ if the following situation occurs $ \pi=[\cdots\rightarrow u\leftarrow\cdots]$. 

We say that a path  $\pi$ from $u$ to $v$ in a DAG
$\dag$ is  \emph{connecting} relative to $S$, if all colliders on $\pi$ are in the ancestral set $\An(S)$, and all non-colliders are outside $S$. A path that is not connecting relative to $S$ is said to be  {\em
blocked\/}  by $S$. 
We then define an independence model $\dse$ relative to a directed graph $\dag$ as follows: \bde\label{def:dsep}\rm For three disjoint subsets $A$, $B$, and $S$  of the vertex set $V$ of a graph $\graf= (V,E)$ we say that $A$ and $B$ are \emph{$\dag$-separated}
by $S$ if all paths from $A$ to $B$ are
blocked by $S$ and we then write $A\dse B\cd S$.
\ede

\bexam\label{ex:dsep}
Consider the network in the figure below, only slightly more complex than in Example~\ref{ex:maxlin}:
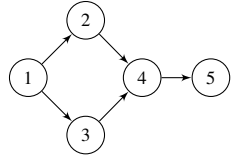
\begin{center}
\begin{tikzpicture}[node distance = 4mm and 4mm, minimum width = 5mm]
    %% nodes      
    \begin{scope}
      \tikzstyle{every node} = [shape = circle, 
      font = \scriptsize,
      minimum height = 4mm,
      inner sep = 0pt,
      draw = black, 
      fill = white, 
      anchor = center], 
      text centered] 
      \node(a) at (0,0) {$1$};
      \node(b) [above right = of a] {$2$};
      \node(c) [below right = of a] {$3$}; 
      \node(d) [right =10mm of a]{$4$};
      \node(e) [right = of d]{$5$};
    \end{scope}  

    %% directed edges
    \begin{scope}[->, > = latex']
		\draw (a) -- (b);
    \draw (a) -- (c);
   \draw (c) -- (d);
	\draw (b) -- (d);
	\draw (d) -- (e);

    \end{scope}

    \end{tikzpicture}
    \end{center}
    We have $2\dse 3\cd 1$ since the path $2\leftarrow 1\to 3$ is blocked as the non-collider $1$ is in $S=\{1\}$ whereas the path $2\to 4\to 3$ is blocked because the collider $4$ is not an ancestor of $S=\{1\}$; on the other hand it holds that  $\neg(2\dse 3\cd \{1,5\})$ since now the second path is rendered active as the collider $4$ is in $\An(\{1,5\})$. 
    \eexam
Note that this definition in a natural way extends that of $\gse$ for an undirected graph, as an undirected graph does not have colliders. The independence model $\dse$ also satisfies the graphoid axioms, see e.g.\ \citet{lauritzen:sadeghi:17}.

There is an alternative method for checking $\dag$-separation in terms of standard separation in a suitable undirected graph, associated with the query. 
More precisely we say that $A$ is \emph{$m$-separated} from $B$ by  $S$ and we write $A\mse B\cd S$ if $S$ separates $A$ from $B$ in
the moral graph $(\dag_{\An(A\cup B\cup S)})^{m}$.  We then have:

\begin{proposition}\label{sepcrit}
Let $A$, $B$ and $S$ be disjoint subsets of the nodes of a directed 
acyclic graph $\graf$. Then $A\dse B\cd S  \iff A\mse B \cd S$. 
\end{proposition}
For a proof, see \citet{richardson:03}, amending an inaccuracy in \citet{Lauritzen1990}.

\bexam\label{ex:dsep2}To illustrate the alternative procedure, we 
again consider the network in Example~\ref{ex:dsep}.

    If we wish to check whether $ 2\dse 3 \cd 1$ we consider the subgraph induced by the ancestral set of $\{1,2,3\}$ and moralize to obtain the graph to the left in the figure below. Since $1$ is  a separator in this graph, we conclude that $ 2\dse 3 \cd 1$.
    
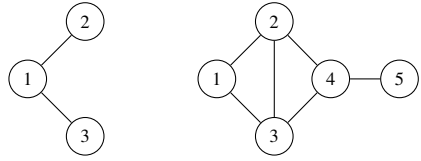
\begin{center}
\begin{tikzpicture}[node distance = 4mm and 4mm, minimum width = 5mm]
    %% nodes      
    \begin{scope}
      \tikzstyle{every node} = [shape = circle, 
      font = \scriptsize,
      minimum height = 4mm,
      inner sep = 0pt,
      draw = black, 
      fill = white, 
      anchor = center], 
      text centered] 
      \node(a) at (0,0) {$1$};
      \node(b) [above right = of a] {$2$};
      \node(c) [below right = of a] {$3$}; 
      \node(d) [right =10mm of a]{$4$};
     \node(e)[right = of d]{$5$};
     \node(a1)[left = 20mm of a]{$1$};
     \node(b1)[above right = of a1]{$2$};
       \node(c1) [below right = of a1] {$3$}; 
     % \node 
    \end{scope}  

    %% indirected edges
    \begin{scope}%[->, > = latex']
		\draw (a) -- (b);
    \draw (a) -- (c);
   \draw (c) -- (d);
	\draw (b) -- (d);
	\draw (d) -- (e);
	\draw (b) -- (c);
	\draw (a1) -- (b1);
	\draw (a1) -- (c1);
    \end{scope}

    \end{tikzpicture}
    \end{center}
    On the other hand, if the query is  whether $2\dse 3\cd \{1,5\}$ we have $\An(\{1,5\})=V$ and thus the relevant moral graph is given to the right in the figure above; in this graph, $2$ and $3$ are not separated by  $\{1,5\}$ so we conclude $\neg(2\dse 3\cd \{1,5\})$.
    \eexam

\subsection{Markov properties of Bayesian networks}\label{sec:bn}

It follows directly from the construction of a Bayesian network, that the joint distribution $P$ satisfies the \emph{well-ordered Markov property} (O) w.r.t.\ $\dag$ if for some well-ordering of $V$,  every variable is conditionally independent of its predecessors given its parents
\[v\cip \pr(v)\cd \pa(v)\]
for all $v\in V=\{1,\ldots, d\}$.

We further say that  $P$ obeys the {\em
local Markov property\/}\index{Markov property!directed!local}
(L) w.r.t.\ $\dag$ if every variable is conditionally independent of its 
non-descendants, given its parents: 
 \[v\cip
(\nd(v)\setminus\parents(v))\cd\parents(v).\]  And, finally, $P$ satisfies the \emph{global Markov property} (G) w.r.t.\ $\dag$ if 
\[A\dse B\cd C \implies A\ci B\cd C.\]

\bexam\label{ex:markovprops}
Consider the network in the figure below:
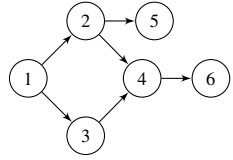
\begin{center}
\begin{tikzpicture}[node distance = 4mm and 4mm, minimum width = 5mm]
    %% nodes      
    \begin{scope}
      \tikzstyle{every node} = [shape = circle, 
      font = \scriptsize,
      minimum height = 4mm,
      inner sep = 0pt,
      draw = black, 
      fill = white, 
      anchor = center], 
      text centered] 
      \node(a) at (0,0) {$1$};
      \node(b) [above right = of a] {$2$};
      \node(c) [below right = of a] {$3$}; 
      \node(d) [right =10mm of a]{$4$};
      \node(e) [right = of d]{$6$};
      \node (f) [right = of b] {$5$};
    \end{scope}  

    %% directed edges
    \begin{scope}[->, > = latex']
		\draw (a) -- (b);
    \draw (a) -- (c);
   \draw (c) -- (d);
	\draw (b) -- (d);
	\draw (d) -- (e);
	\draw (b) -- (f);

    \end{scope}

    \end{tikzpicture}
    \end{center}
 The numbering of the nodes here constitute a well-ordering so, for example, (O) implies $5\ci \{1,3,4\}\cd 2$, whereas the local Markov property (L) implies $5\ci \{1,3,4,6\}\cd 2$; the global Markov property implies, for example, $5\ci \{1,6\}\cd 4$. 
    \eexam
In the case of undirected graphs, the local and global Markov properties are  different \citep[Section 3.2]{Lauritzen1996},  but here we have
\bthe Let $\dag$ be a directed acyclic graph with $V=\{1,\ldots,d\}$ well-ordered and $P$  a probability distribution on $\calx=\times_{v\in V} \calx_v$. Then we have
\[\mbox{\rm (O)}\iff  \mbox{\rm (L)} \iff \mbox{\rm (G)}.\]
In words, if $P$ satisfies any of these Markov properties, it satisfies all of them.
\ethe
\bproof This fact is established in \cite[Corollary 2]{Lauritzen1990} for any semi-graphoid independence model $\sg$. 
\eproof
Note that in particular it is true that \emph{if $P$ satisfies (O) w.r.t.\ one well-ordering, it satisfies (O) w.r.t.\ all well-orderings.}

The global Markov property gives a sufficient condition for conditional independence in terms of  $\dag$-separation. Another central concept is that of faithfulness, formally defined below 
\bde\label{faith}
 A probability distribution $P$ on $\calx=\times_{v\in V} \calx_v$ is said to be \emph{faithful} to a DAG $\dag$ if \[A\dse B\cd C\iff A\ci_P B\cd C.\]
In other words,  if $\dag$-separation is also necessary for conditional independence. 
\ede
Generally, most probability distributions are faithful \citep{Meek95uai}, but we shall later see that this is not the case for the special Bayesian networks we study here.

Finally, we need to emphasize that two different DAGs can define exactly the same independence model.  
Consider two graphs $\dag_1$ and $\dag_2$ as well as their associated independence models $\perp_{\dag_1}$ and $\perp_{\dag_2}$. It may well happen that even though the graphs are different, their independence models might be identical, see for example Figure~\ref{fig:cipequiv} below.
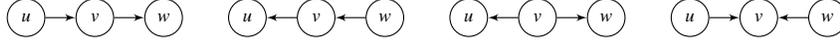
\begin{figure}[htb]
\centering
\begin{tikzpicture}[node distance = 4mm and 4mm, minimum width = 4mm]
    %% nodes      
    \begin{scope}
      \tikzstyle{every node} = [shape = circle, 
      font = \scriptsize,
      minimum height = 5mm,
      inner sep = 0pt,
      draw = black, 
      fill = white, 
      anchor = center], 
      text centered] 
      \node(a) at (0,0) {$u $};
      \node(b) [right = of a] {$v$};
      \node(c) [right = of b] {$w$};
       
      \node(d) [right = 6mm of c] {$u$}; 
      \node(e) [right = of d] {$v$};
      \node(f) [right = of e] {$w$};
      
     \node(g) [right = 6mm of f] {$u$}; 
      \node(h) [right = of g] {$v$};
      \node(i) [right = of h] {$w$};
      
     \node(j) [right = 6mm of i] {$u$}; 
      \node(k) [right = of j] {$v$};
      \node(l) [right = of k] {$w$};
    \end{scope}  
    %% directed edges
    \begin{scope}[->, > = latex']
		\draw (a) -- (b);
    \draw (b) -- (c);
   \draw (f) -- (e);
	\draw (e) -- (d);
	\draw (h) -- (g);
	\draw (h) -- (i);
	\draw (j) -- (k);
	\draw (l) -- (k);
    \end{scope}
  \end{tikzpicture}
 \caption{\label{fig:cipequiv} The DAGs to the left of the figure are Markov equivalent; the only non-trivial element of their independence models is $u\dse w\cd v$.  The DAG to the right in the figure has a different independence model, since there $u\dse w$.} 
\end{figure}

Here all independence models are the same although the graphs are different. This also means that any probability distribution $P$ which satisfies the global Markov property for any of them, automatically satisfies the global Markov property for all of them.
We formally define
\bde Two DAGs $\dag_1$ and $\dag_2$ are \emph{Markov equivalent}\index{Markov equivalence} if and only if their independence models coincide, i.e.\ if 
$A\perp_{\dag_1} B\cd C \iff A\perp_{\dag_2}  B\cd C$.
\ede
The following result was shown by \citet{frydenberg:90} and \citet{verma:pearl:90} and gives a necessary and sufficient condition for two DAGs to be Markov equivalent.
\bthe\label{thm:markoveqiuv}Two directed acyclic graphs $\dag_1= (V,E_1)$ and $\dag_2=(V,E_2)$ are Markov equivalent if and only if they have the same skeleton $\ske(\dag_1)=\ske(\dag_2)$ and the same unshielded colliders.
\ethe 
Here the \emph{skeleton} $\ske(\dag)$ of a DAG $\dag$ is the undirected graph with $u\sim v$ in $\ske(\dag)$ if $u\sim v$ in $\dag$, and an \emph{unshielded collider} is a triple $u\to w\leftarrow v$ with $u\not\sim v$.

\section{Recursive max-linear structural equation models}\label{sec:rec}

We shall be interested in Bayesian networks defined through structural equation systems (\ref{eq:structure}) where the functions $g_v$ are \emph{max-linear,} i.e.\ the additions in (\ref{eq:linear}) are replaced with the operation of forming the maximum.

Henceforth we assume that the vertex set of our DAG $\dag=(V,E)$ is well-ordered  so $V=\{1,\ldots,d\}$ and assume a data generating mechanism specified via a \emph{recursive max-linear structural equation model}, which has representation
\begin{align} \label{ml-sem}
X_v =\bigvee\limits_{u \in \pa(v)} c_{vu} X_u \vee c_{vv} Z_v,\quad v=1,\dots,d,
\end{align}
where $Z_1, \dots, Z_d$ are independent and identically distributed with a continuous distribution having support $\R_+=(0,\infty)$, and $c_{vu}>0$, $u\in\pa(v)$, $c_{vv}$ are \emph{structural coefficients} in the equations or \emph{edge weights} for the associated DAG $\dag$.

Following \citet{GK1} we say this is a \emph{recursive max-linear model}. Note that our use of indices for edge weights here is the opposite of that used in \cite{GK1}. 

For simplicity we assume throughout the rest of the paper that $c_{vv}=1$ for all $v\in V$.
For a path $\pi=[u=k_0 \rightarrow k_1\rightarrow \dots \rightarrow  k_n=v]$ of length $n$ from $u$ to $v$, we define the quantities
\begin{align} \label{bs}
d_{vu}(\pi) :=  \prod_{l=0}^{n-1} c_{k_{l+1} k_{l}}\quad\mbox{ and }\quad
b_{vu}:=\bigvee_{\pi\in \Pi_{uv}} d_{vu} (\pi),
\end{align}
where $\Pi_{uv}$ denotes all paths from $u$ to $v$.
In summary, we define
\begin{align}\label{bs-max}
b_{vu}=\bigvee\limits_{\pi \in \Pi_{uv}} d_{vu}(\pi)\mbox{ for $u \in {\an}(v)$};\,\, b_{vv}= c_{vv}=1;\,\,\mbox{$b_{vu}=0$ for $u\in {V\setminus \An(v)}$},
\end{align}  
where  $\An(v)=\an(v)\cup \{v\}$ is the smallest ancestral set containing vertex $v$.
We then arrange these coefficients in the {\em max-linear coefficient matrix\/} $B = (b_{vu})_{d\times d}$ and find
\begin{align} \label{ml-noise}
X_v=\bigvee\limits_{u \in \An(v)} b_{vu}Z_u, \quad v=1,\dots,d.
\end{align}
This equation represents $X$ as a \emph{max-linear model} as defined for instance in \citet{Wang2011}. 

For two non-negative matrices $F$ and $G$, where the number $n$ of columns in $F$ is equal to the number of rows in $G$ we introduce the product $\odot$ as
\begin{align}\label{odot} 
(F\odot G)_{vu}=\Big(\bigvee\limits_{k=1}^n f_{vk}g_{ku}\Big). 
\end{align}
If we collect the noise variables into the column vector $Z=(Z_1,\ldots,Z_d)^\prime$, the representation \eqref{ml-noise}  of $X$ can then  be written as
\begin{align*} 
X=B \odot Z = \big(\bigvee_{u=1}^d b_{vu} Z_j, i=1,\ldots,d\big) = \big(\bigvee_{u\in\An(v)} b_{vu} Z_j, i=1,\ldots,d\big).
\end{align*}

Given the \DAG\  $\dag$ and the edge weights $c_{ik}$ with $c_{ii}=1$ for all $i=1,\dots,d$,
the max-linear coefficient matrix $B$ can be found by iterating the weighted adjacency matrix $\coeff=(c_{vu} \1_{\Pa(v)}(u))_{d\times d}$ of $\dag$ using this matrix multiplication;  here $ \1_{\Pa(v)}$ denotes the indicator function of $\Pa(v)=\pa(v)\cup\{v\}$) :
\beam\label{BC}
B=  \bigvee_{k=0}^{d-1}\coeff^{\odot k} = C^{\odot(d-1)},
\eeam
cf. \citet{Butkovic}, Lemma~1.4.1.
For more details see \cite{GK1}, Theorem~2.4.

By \eqref{bs-max} the max-linear coefficient $b_{vu}$ of $X$ is different from zero if and only if $u\in \An(v)$.
This information is contained in the  {\em reachability matrix\/}  $R=(r_{vu})_{d \times d}$ of $\dag$, which has entries
\begin{align*}
r_{vu}:=\begin{cases}
1, & \text{if there is a path from $u$ to $v$, or if $u=v$},\\
0, & \text{otherwise}. 
\end{cases}
\end{align*}
If the $vu$-th entry of $R$ is equal to one, then $v$ {\em is reachable from} $u$. 
In the context of a \DAG\ $\dag$ with its reachability matrix $R$ and a recursive max-linear model $X$ on $\dag$ with max-linear coefficient matrix $B$  it will be useful to keep the following in mind.

\brem\label{maxlinear_struc} 
Let  $\dag$ be a \DAG\ with reachability matrix $R$.\\[1mm]
(i) \, The max-linear coefficient matrix $B$ is a weighted reachability matrix of  $\dag$; i.e., $R=\sgn(B)$.\\[1mm]
(ii) \, 
Since $V$ is assumed well-ordered,
$B$ and $R$ are lower triangular matrices.
\erem

From \eqref{bs-max} and \eqref{ml-noise} we conclude that a path $\pi$ from $u$ to $v$, whose weight $d_{vu}(\pi)$ is strictly less than $b_{vu}$ does not have any influence on $X_i$.  
For $v\in V$ and $u\in \an(v)$ we call a path $\pi$ from $u$ to $v$ \textit{max-weighted}, if $b_{vu}=d_{vu}(\pi)$, and investigate its relevance for the recursive max-linear model in further detail. 

Firstly we note that we can remove an edge from $\dag$ which is not part of a max-weighted path without changing the distribution of $X$. The DAG obtained in this way is termed the \emph{minimum max-linear \DAG} $\dag^B$. 
In the special case where $\dag$ is a polytree, all paths are necessarily max-weighted and we clearly have
\bpr If $\dag$ is a polytree, it holds that $\dag^B=\dag$.
\epr

The following result describes exactly all DAGs and edge weights possible for a given max-linear coefficient matrix.
Recall that we have set $c_{vv}=1$.

\bthe\label{le1:CMgraph}[\citet{GK1}, Theorem~5.4]\\
Let $X$ be given by a recursive max-linear structural equation system with coefficient matrix $B$.  
Let further $\dag^B=(V,E^B)$ be the minimum max-linear \DAG\ of $X$ and $\pa^B(v)$ the parents of $v$ in  $\dag^B$. 
\begin{enumerate}
\item[(a)]  
$\dag^B$ is  the \DAG\  with the minimum number of edges such that $X$ satisfies \eqref{ml-sem}. 
The weights in \eqref{ml-sem} are uniquely given by $c_{vv}=b_{vv}$ and $c_{vs}=b_{vs}$  for $v\in V$ and $s\in\pa^B(v)$. 
\item[(b)] Every  \DAG\  with vertex set $V$ that has at least the edges of $\dag^B$ and the same reachability matrix as $\dag^B$ represents $X$ in the sense of \eqref{ml-sem} with weights satisfying
 \begin{align*}
 c_{vv}=b_{vv}, \, c_{vs}=b_{vs}\text{ for $s\in  \pa^B(v)$}, \text{ and } c_{vs}\in \left(0,b_{vs}\right)\text{ for $s\in \pa(v)\setminus \pa^B(v)$}.  
 \end{align*}
There are no further \DAG s and   weights   such that  $X$ has representation \eqref{ml-sem}. 
\end{enumerate}
\ethe

In general, recursive max-linear models are not faithful to their DAG, not even if $\dag=\dag^B$, see Remark 3.9 (ii) in \cite{GK1}. This is illustrated in Example~\ref{ex:ml} below.

\bexam\label{ex:ml}[Example 3.8 of \cite{GK1} and continuation of Example~\ref{ex:maxlin}:] 
We note that the paths $[1\to 2]$, $[1\to 3]$, $[2\to 4]$, and $[3\to 4]$ are max-weighted as they are the only directed paths between their endpoints. It therefore holds that $\dag^B=\dag$ since they are the unique max-weighted paths. 
Still, the distribution determined by this recursive system is never faithful to $\dag$, as we shall see below.

Concerning the paths from node $1$ to $4$ we have three situations:
\begin{align*}
c_{42}c_{21}=c_{43}c_{31}, \quad c_{42}c_{21}>c_{43}c_{31}, \quad \text{or}\quad c_{42}c_{21}<c_{43}c_{31}.
\end{align*}
In the first situation, both paths from $1$ to $4$, $[1\to 2 \to 4]$ and $[1\to 3 \to 4]$, are max-weighted whereas in the other situations only one of them is.   

If the path $[1\to2\to 4]$ is max-weighted, we can consider the subdag $\tilde\dag$ obtained from $\dag$ by removing the edge $1\to3$:
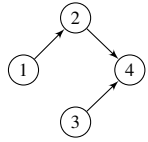
\begin{center}
\begin{tikzpicture}[node distance = 4mm and 4mm, minimum width = 4mm]
    %% nodes      
    \begin{scope}
      \tikzstyle{every node} = [shape = circle, 
      font = \scriptsize,
      minimum height = 4mm,
      inner sep = 0pt,
      draw = black, 
      fill = white, 
      anchor = center], 
      text centered] 
      \node(a) at (0,0) {$1$};
      \node(b) [above right = of a] {$2$};
      \node(c) [below right = of a] {$3$}; 
      \node(d) [right =10mm of a]{$4$};
    \end{scope}  

    %% directed edges
    \begin{scope}[->, > = latex']
		\draw (a) -- (b);
    %\draw (a) -- (c);
   \draw (c) -- (d);
	\draw (b) -- (d);
    \end{scope}

    \end{tikzpicture}
    \end{center}
In other words, we are changing the edge weights by letting $\tilde c_{31}=0$, keeping the other edge weights unchanged. The new max-linear coefficient matrix becomes 
\begin{align*}
\tilde B = 
\begin{pmatrix}
1 &  0    & 0 &  0   \\
c_{21}  & 1 &   0    &  0 \\
0   &    0   &   1&   0  \\
c_{42}c_{21} & c_{42} & c_{43} &  1
\end{pmatrix}
\end{align*}
where we have exploited that $c_{ii}=1$.
The max-linear coefficient matrix for the marginal distribution of $(X_1,X_2,X_4)$ is obtained by ignoring the third row 
and since only entries in the third row have changed, we see that  $(X_1,X_2,X_4)$ has the same joint distribution in the model determined by $\dag$ as it has in the model determined by $\tilde\dag$. 

But as we clearly have $1\perp_{\tilde\dag}4\cd 2$, we conclude that $X_1\cip X_4\cd X_2$ in the model determined by $\tilde \dag$ and hence also by $\dag$. But since $\neg(1\dse 4\cd 2)$, the distribution is not faithful to $\dag$.   

If $[1\to3\to 4]$ is also max-weighted, the similar argument yields $X_1\cip X_4\cd X_3$, so the distribution is \emph{not faithful to $\dag$} for any allocation of edge weights.
\eexam

We note that
 \cite{GK1} suggest in their  Remark~3.9(i)  that additional conditional independence relations that are valid for a given DAG can be revealed by considering a system of submodels determined by appropriate subgraphs, but here we refrain from giving a complete description of all valid conditional independence relations.

\section{Statistical properties}\label{sec:stats}

The statistical theory of recursive max-linear models is challenging because standard assumptions for smooth statistical models are not satisfied. For example, if we for a given DAG $\dag$ consider the family $\cal P$ of distributions with coefficients adapted to $\dag$, this family is not dominated by any measure on the space of observations, so standard likelihood theory does not apply.  On the other hand, as we shall see, estimation of coefficients and identification of the network structure for recursive max-linear models can be made in a simple fashion and procedures are more efficient than usual in that estimates of coefficients and structures converge at exponential rates to the true values. Here we shall give a summary of the most important findings in \citet{GKL}.

Throughout the following we consider a sample $\mathbf{x}=(X^1=x^1,\ldots,X^n=x^n)$ from a distribution $P$ given by the recursive max-linear model (\ref{ml-sem}).

\subsection{Estimation of coefficients}

We first consider the situation where the DAG $\dag=(V,E)$ and for the sake of simplicity we assume the distribution of noise variables $Z_v,v\in V,$ is completely known, the coefficients $c_{vv}$ are all equal to one, whereas the edge weights $\coeff=\{c_{vu}, u\in \parents(v), v\in V\}$ are all strictly positive, but otherwise unknown. We let $\Coeff$ denote the set of all possible coefficients and $P_\coeff$ denote the distribution of $X$ determined by the  corresponding recursive model \eqref{ml-sem}. 

The family ${\cal P}=P_\coeff, \coeff\in \Coeff,$ is not dominated by any fixed $\sigma$-finite measure $\mu$ on $\calx$, as the support of $P_\coeff$ varies strongly with the coefficients; more precisely, the distributions have disjoint atomic components. This is a disadvantage in the sense that we cannot define a standard likelihood function; but, as we shall see, an advantage since these atomic components help identifying  $P_\coeff$ from a given sample. We illustrate this by a simple example.

\bexam\label{ex:simpledag}[Estimation from the atoms]\\
Consider the simple DAG $1\to 2$ with just two nodes and a single directed edge, and let $c=c_{21}$ be the corresponding coefficient.

Then $P_c$ has support on the cone given as $x_2\geq cx_1\geq 0$ and the line $A_c=\{x_2=cx_1\}$ is an atom for $P_c$ because $P_c(A_c)= P(Z_2\leq cX_1)= P(Z_2\leq cZ_1)>0$ (cf. Remark~\ref{rem}(ii)).

Still, since then $\{c\}$ is the only atom in $P_c$ for $Y=X_2/X_1$, the sample will for large $n$ with high probability have repeated values of $Y$ and $c$ will be the only  value that is repeated. 
 In other words, $\hat c = \min\{y^\nu =x_2^\nu/x^\nu_1, \nu=1, \ldots, n\}$ will be exactly equal to the true parameter with high probability.
A similar estimator has been considered by \citet{Davis1989} in a time-series framework.
\eexam

Although most likelihood theory is concerned with dominated families, \citet{KW1956} considered the non-dominated case.  Their formulation has been used rarely --- an exception being 
\citet{Johansen1978}; see also \citet{scholz:80} and \citet{gill:etal:89}, for example. This formulation turns out to be exactly what we need to discuss estimation of $\coeff$ in a formal way. 

For two probability measures $P$ and $Q$ on a measurable space $(\spa, \mathbb{E})$, we define the \emph{generalized likelihood ratio} $\rho_x(P,Q)$ at the observation $x$ as 
\begin{equation}
\rho_{x}(P,Q)= \frac{\dd P}{\dd (P+Q)}(x)
\end{equation}
where ${\dd P}/{\dd (P+Q)}$ is the density of $P$ w.r.t.\ $P+Q$; the density always exists as, clearly, $P(A)+Q(A)=0\implies P(A)=0$ so $P$ is absolutely continuous w.r.t.\ $P+Q$.

The idea here is that if $\rho_x(P,Q) >\rho_x(Q,P)$, then $P$ is a more likely explanation of $x$ than $Q$. We note in particular that if $P$ and $Q$ have densities $f$ and $g$ w.r.t.\ a $\sigma$-finite measure $\mu$, we have $\rho_x(P,Q)= f(x)/\{f(x)+g(x)\}$ so then $\rho_x(P,Q)>\rho_x(Q,P)$ if and only if $f(x)>g(x)$. Hence $\rho_x$ extends the standard likelihood ratio in a natural way.

Clearly, the generalized likelihood ratio suffers from the same problem as the usual likelihood ratio: the densities are only defined almost surely, so can be changed on $P+Q$-null sets; therefore, a version of ${\dd P}/{\dd (P+Q)}$ must be chosen independently of the  observation $x$.

Next we say that if $\calp$ is a family of probability distributions, $\hat P$ is a \emph{generalized maximum likelihood estimate} (GMLE) of $P$ based on $x\in \supp(\hat P)$ if
\[\rho_x(\hat P,Q)\geq \rho_x(Q,\hat P)\mbox{ for all $Q\in \calp$,}\]
i.e.\ if $\hat P$ explains $x$ at least as well as any other member of $\calp$.

\bexam\label{ex:simpledag_ext}[Continuation of Example~\ref{ex:simpledag}: GMLE]\\
We illustrate use of  the generalized maximum likelihood ratio for the model described in Example~\ref{ex:simpledag}. To identify the density, we consider two  values  $c> c^*$ where we have
\[\rho_x(c,c^*)=\frac{\dd P_c}{\dd (P_c+P_{c^*})}(x_1,x_2)=
\begin{cases}
1/2& \text{ for } x_2>cx_1\\
1 & \text{ for } x_2=cx_1\\
0 & \text { for } x_2< cx_1
\end{cases}
\] 
and
\[\rho_x(c^*,c)=\frac{\dd P_{c^*}}{\dd (P_c+P_{c^*})}(x_1,x_2)=
\begin{cases}
1/2& \text{ for } x_2>cx_1\\
0 & \text{ for } x_2 =cx_1\\
1& \text{ for } cx_1>x_2\geq c^*x_1\\
0 & \text { for } x_2< c^*x_1.
\end{cases}\]
If $c=c^*$ we may let
$$\rho_x(c,c)=\rho_x(c,c^*)=\rho_x(c^*,c)=\frac12\1_{\{x_2\geq cx_1\}}.$$
Thus, if we consider a full sample, let $\hat c = \min
\{y^\nu =x_2^\nu/x^\nu_1, \nu=1, \ldots, n\}$ 
and $n_+(c, \mathbf{x})=\#\{\nu: y^\nu> c\}$, we get: 
\[\rho_\mathbf{x}(\hat c, c)=\prod_{\nu=1}^n\rho_{x^\nu}(\hat c,c)=
\begin{cases}0& \text{ if } c> \hat c \text{ and } c\in \{y^\nu, \nu=1, \ldots, n\}\\
2^{-n_+(c, \mathbf{x})}&\text { if }  c> \hat c \text{ and } c\not\in \{y^\nu, \nu=1, \ldots, n\}\\ 
2^{-n} &\text{ if } c=\hat c\\
2^{-n_+(\hat c, \mathbf{x})} &\text{ if }  c< \hat c,\end{cases}\]
whereas 
\[\rho_\mathbf{x}( c,\hat c)=\prod_{\nu=1}^n\rho_{x^\nu}(c,\hat c)=
\begin{cases}0& \text{ if } c> \hat c\\
2^{-n} &\text{ if } c=\hat c\\
0&\text{ if }  c< \hat c.\end{cases}\]
Clearly,  $\rho_\mathbf{x}(\hat c, c)\geq \rho_\mathbf{x}( c,\hat c)$ showing that $\hat c$ is the unique GMLE of $c$.
\eexam

Indeed, \emph{it holds in general for a recursive max-linear model that} 
$$\hat c_{ij}= \bigwedge_{\nu=1}^n \, \frac{x^\nu_i}{x^\nu_j}, \quad i\in V, j\in \parents(i)$$
\emph{is a GMLE of the edge weights.}
We refer to \cite{GKL} for further details but should point out that in the general case, the GMLE is not unique. Since the distribution of $X$ only depends on the edge weights through the max-linear coefficient matrix $B$, only $B$ is uniquely estimable from a sample. We clearly have by \eqref{BC} for the GMLE that 
\[\hat B= B(\hat \coeff) = \bigvee_{k=0}^{d-1} \hat \coeff^{\odot k} = \hat C^{\odot(d-1)}.\]
An alternative estimate of the max-linear coefficient matrix is given as
\[\tilde b_{ij}=\bigwedge_{\nu=1}^n \, \frac{x^\nu_i}{x^\nu_j}, \quad i\in V, j\in \an(i).\]
Although this estimate is also sensible and asymptotically consistent, it is less efficient than the GMLE as $X^\nu_i/X^\nu_j$ only attends its minimum value when all noise variables on the path from $j$ to $i$ are smaller than $b_{ij}X^\nu_j$ for the same $\nu$, whereas the minima for the $X^\nu_{v}/X^\nu_{u}$ on the path from $j$ to $i$ can be attained for different $\nu$s. 

\subsection{Identification of structure}
General methods for identifying the structure  of DAG $\dag$ from a sample are often based on an assumption of faithfulness, so that observed conditional independence relations can be translated back to the structure of the DAG since then any observed conditional independence must correspond to a separation in $\dag$, see for example \citet{SGS}.  
Also, as noted in Theorem~\ref{thm:markoveqiuv}, two DAGs can be different but still Markov equivalent and thus any method based on observed direct conditional independence relations cannot distinguish between DAGs that are Markov equivalent.

As shown in Example~\ref{ex:ml}, faithfulness is violated  for max-linear Bayesian networks whenever $\dag$ is not a polytree. However, as we shall see below, the minimal DAG $\dag^B$ of a max-linear Bayesian network can still be completely recovered from observations.

This fact conforms with recent developments where the recursive linear structural equation systems have been shown to be completely identifiable if the errors follow a non-Gaussian distribution (\citet{Shimizu2006}) and it has been shown that the faithfulness assumption can be considerably weakened also in other situations (\citet{SpirtesZhang,BP2014}).

To explain why the structure $\dag^B$ is identifiable, we consider the statistics
\[Y_{ij}=X_i/X_j, \quad i,j\in V\]
and note that  $Y_{ij}$ has support $[b_{ij},\infty)$ and an atom in $b_{ij}$ if and only if $j\in \an(i)$. Using this property one can show that the following estimate $\check B$  eventually identifies the max-linear coefficient matrix $B$. 
\[\check b_{ij} = \begin{cases}\bigwedge\limits_{\nu=1}^n y^\nu_{ij} & \text{ if  minimum value is attained at least twice in the sample, }\\
0& \text{ otherwise.}
\end{cases}\]
Then $\dag^B$ is identifiable from $B$; we refer the reader to \cite{GKL} for further details.

\section{Conclusion}
We have reviewed basic elements of Bayesian networks based on recursive max-linear structural equations and some of their statistical properties.
We conclude this article by pointing out some natural extensions of this work that we hope to address in the future.  

Firstly, it would be of interest to have a simple and complete description of all independence properties which hold for a distribution determined by a recursive max-linear equation system, i.e.\ a global Markov property for max-linear Bayesian networks. 

Secondly, it appears that a consequent use of algebraic theory; see e.g.\ \citet{Butkovic}, based on properties of the max-times semiring $\mathbb{S}=([0,\infty], \vee, \cdot)$ would be able to simplify the theory of these models.  

Finally, we should emphasize that the models heuristically can be seen as limiting cases of standard linear recursive models where error distributions have heavy tails and therefore the maximal element of any sum will almost completely dominate the sum; a rigorous study of this limiting process will enhance the understanding of this class of models.

\section*{Acknowledgements}The authors have benefited from discussions with Nadine Gissibl and financial support from the Alexander von Humboldt Stiftung.

\end{document}